\documentclass[aps,pra,twocolumn,superscriptaddress,amssymb,showpacs]{revtex4-1}
\usepackage{graphicx}
\usepackage{epstopdf}

\begin{document}

\title{Generation of Fock states and qubits in periodically pulsed nonlinear oscillator}

\author{T.~V.~Gevorgyan}
\email[]{t\_gevorgyan@ysu.am}
\affiliation{Institute for Physical Researches, National Academy
of Sciences,\\Ashtarak-2, 0203, Ashtarak, Armenia}

\author{A.~R.~Shahinyan}
\email[]{anna\_shahinyan@ysu.am}
\affiliation{Yerevan State University, Alex Manoogian 1, 0025,
Yerevan, Armenia}

\author{G.~Yu.~Kryuchkyan}
\email[]{kryuchkyan@ysu.am}
\affiliation{Institute for Physical Researches,
National Academy of Sciences,\\Ashtarak-2, 0203, Ashtarak,
Armenia}\affiliation{Yerevan State University, Alex Manoogian 1, 0025,
Yerevan, Armenia}

\begin{abstract}
We demonstrate a quantum regime of dissipative nonlinear oscillators where the
creation of Fock states as well as the superpositions of Fock states are
realized for time-intervals exceeding the characteristic decoherence time.
The preparation of quantum states is conditioned by strong Kerr nonlinearity as
well as by excitation of resolved lower oscillatory energy levels with a
specific train of Gaussian pulses. This provides practical signatures to look
for in experiments with cooled nonlinear oscillators.

\end{abstract}


\maketitle

\section{Introduction}

The preparation and use of Fock states, that have definite numbers of energy
quanta, and various superpositions of Fock states form the basis of quantum
computation and communications \cite{Niel}. Quantum dynamics of an oscillator
is naturally described by Fock states, however, these states are hard to create
in experiments. The reason is that excitations of oscillatory systems usually
lead to the production of
thermal states or coherent states but not quantum Fock states.
Nevertheless, quantum oscillatory states can be prepared and can be manipulated by
coupling oscillators to atomic systems. In this way, a classical pulse applied
to the atomic states creates a quantum state that can subsequently be
transferred to the harmonic oscillator excited in a coherent state. The
systematic procedure has been proposed in Ref. \cite{law} and has been demonstrated
for deterministic preparation of mechanical oscillatory Fock states with
trapped ions \cite{mek} and in cavity QEDs with Rydberg atoms \cite{varc}.
Most recently the analogous procedure has been applied in solid-state
circuit QED for deterministic preparation of photon number states in a
resonator by interposing a highly nonlinear Josephson phase qubit between a
superconducting resonator \cite{hof}.

In this paper, we show that the production of Fock states and their various two-state superpositions or qubits can also be realized
for an anharmonic oscillator without any interactions
with atomic and spin-1/2 systems. For this goal we consider nonlinear
dissipative oscillator the (NDO) under sequence of classical Gaussian
pulses separated by time intervals. The nonlinearity effects in the NDO should be
enough to break the equidistant of oscillatory energy levels. In this case, the
oscillatory energy levels are well resolved, and selection is possible near the
resonant excitation for them. It is demonstrated that the 
creation of oscillatory nonclassical states can be
realized in a periodically pulsed oscillator with complete consideration of decoherence effects and in a regime of strong Kerr nonlinearity.

It is well assessed, that in a free decoherence regime an anharmonic oscillator
leads to Schr\"{o}dinger cat states which are destroyed due to dissipation and
decoherence. This situation of vanishing quantum superposition states of the
NDO in an over transient dissipative operational regime has been strongly
demonstrated within the framework of the Wigner function that visualizes
quantum interference as negative values in the phase space. Really
surprisingly analytical results
for the Wigner functions of the NDO under monochromatic excitation have been
analytically obtained \cite{a33}, \cite{i34} in the steady-state regime that are
positive in all ranges of the phase space. This Wigner function mainly displays
only conventional attributes of the model such as bistability or turning
points. In the extreme quantum regime of the system corresponding to strong
nonlinearities or low damping, the Wigner function displays a rich variety of
phase transition images in the bistable operation regime, qualitatively
different from what one would expect from the corresponding semiclassical
analysis but not squeezing and the other nonclassical effects.

Here, we demonstrate that in the specific pulsed regime of the NDO the
production of the Fock states are realized as well as the superposition of the
Fock states. These states can be created for time intervals exceeding the
characteristic time of decoherence. The corresponding Wigner functions of the 
oscillatory mode show ranges of negative values and gradually deviate from the
Wigner function of an NDO driven by monochromatic driving.

Previously, a driven NDO operated in a quantum regime has become
more important in both fundamental and applied sciences,
particularly, for the implementation of basic quantum optical systems
in the engineering of nonclassical states and quantum logic. In these systems,
the efficiency of the quantum effects requires a high
nonlinearity with respect to dissipation. On the other hand, ground state
cooling is critical for reaching quantum regimes of oscillatory systems. In
this regard, the
largest Kerr nonlinearities were proposed for many physical
systems: cavity quantum electrodynamics-based devices
\cite{Fs}, \cite{Tur}, in electromagnetically induced transparency
\cite{Ima}, and for cooling nanoelectromechanical systems and
nano-opto-mechanical systems based on various oscillators
\cite{craig},\cite{ekinci}. Superconducting devices based on the
nonlinearity of the Josephson junction (JJ) that exhibits a wide
variety of quantum phenomena \cite{21}-\cite{Hosk} offer an unprecedented high
level of nonlinearity and low quantum noise. In some of these devices dynamics
is analogous to those of a quantum particle in an oscillatory anharmonic
potential \cite{claud}, \cite{vijay} and behaves like artificial atoms
\cite{Nori}. Note that the comparison of second-order nonlinearities taking
place for various quantum devices has been recently analyzed in Ref.
\cite{practice} and the NDO for the case of strong nonlinearity has been
considered in \cite{qsch}, \cite{Mih}.

The paper is arranged as follows. In Sec. II, we describe a pulsed NDO and
consider some physical implementations of the model. In Sec. III, we shortly
discuss the NDO under a monochromatic driving for both transient and steady-state
regimes. In Sec.IV, we consider the production of Fock states in the pulsed regime
of the NDO. In Sec. V, we present the results for the construction of quantum
superposition of Fock states on the base of the Rabi oscillations and the
Wigner functions. We summarize our results in Sec. VI.

\section{Model description and implementations}

The Hamiltonian of an anharmonic driven oscillator in the rotating wave approximation reads as:
\begin{equation}
H=\hbar \Delta a^{+}a + \hbar \chi (a^{+}a)^{2} +
\hbar f(t)(\Omega a^{+} + \Omega^{*}a),\label{hamiltonian}
\end{equation}
where time dependent coupling constant $\Omega f(t)$, that is proportional to
the amplitude of the driving field, consists of the Gaussian pulses with 
duration $T$, which are separated by time intervals $\tau$,
\begin{equation}
f(t)=\sum{e^{-(t - t_{0} - n\tau)^{2}/T^{2}}}. \label{driving}
\end{equation}
Here, $a^{+}$, $a$ are the oscillatory creation and annihilation operators,
$\chi$ is the nonlinearity strength, and $\Delta=\omega_{0} -\omega$ is the
detuning between the mean frequency of the driving field and the frequency of
the oscillator.

This model seems experimentally feasible and can be realized in
several physical systems, particularly, for the production of Fock states. In fact, the effective Hamiltonian (\ref{hamiltonian})
describes a nanomechanical oscillator with $a^{+}$ and $a$ raising
and lowering operators related to the position and momentum
operators of a mode quantum motion,
\begin {equation}
x=\sqrt{\frac{\hbar}{2m\omega_0}}(a+a^{+}),\ p=-i\sqrt{2\hbar
m\omega_0}(a-a^{+}) \label{eq:xp},
\end{equation}
where $m$ is the effective mass of the nanomechanical resonator,
$\omega_0$ is the linear resonator frequency, and $\chi$ is 
proportional to the Duffing nonlinearity. It is possible to reach the quantum
regime for such frequencies, i.e., to cool down the temperatures for which
thermal energy is comparable to the energy of oscillatory quanta. Recent
progress in cooling mechanical oscillators to their ground states has been
achieved in Refs. \cite{wilson}-\cite{chan}, (see also Ref. \cite{Tian} and the
references therein). This interest is caused by measuring the quantum effects
\cite{Rips}-\cite{Gaid} and by preparing resonators in states with high purity
to exploit their quantum behavior in future technologies \cite{Wool},
\cite{Rabl}.

One of the variants of
nano-oscillators is based on a double-clamped platinum beam
\cite{N} for which the nonlinearity parameter equals 
$\chi=\hbar/4\sqrt{3}Qma_{c}^{2}$, where $a_c$ is the critical
amplitude at which the resonance amplitude has an infinite slope
as a function of the driving frequency, and $Q$ is the mechanical
quality factor of the resonator. In this case, the giant
nonlinearity with the strength $\chi\cong3.4\cdot10^{-4} s^{-1}$ was realized. Note,
that details of this resonator, including the expression for the
parameter $a_{c}$, are presented in Ref. \cite{amp}.

It is
known that a single light mode propagated in Kerr media with
dielectric constant $\varepsilon$ and $\chi^{(3)}$ susceptibility
is described by the effective Hamiltonian (\ref{hamiltonian}). In
this case the parameter $\chi$ reads as $\chi=\frac{9}{8}
\frac{\hbar\omega^{2}\chi^{(3)}}{\varepsilon^{2}Sd}$, where $S$ is
the cross-sectional area of the beam and $d$ is the envelope width \cite{Drummond},\cite{Mandel}.

The same Hamiltonian
describes a current-biased Josephson junction in the rotating wave
approximation. In this case, creation and annihilation operators
are expressed in the following terms:
\begin{equation}
a=\frac{1}{\sqrt{2}}\left\lbrace
\left(\frac{E_J}{E_C}\right)^{1/4}\phi+i\left(\frac{E_C}{E_J}\right)^{1/4}n\right\rbrace
,\nonumber
\end{equation}
\begin{equation}
\ a^{+}=\frac{1}{\sqrt{2}}\left\lbrace
\left(\frac{E_J}{E_C}\right)^{1/4}\phi-i\left(\frac{E_C}{E_J}\right)^{1/4}n\right\rbrace,
\end{equation}
where $n$ is the number operator of Cooper-pair charges, $\phi$ is the phase difference
between superconducting elements in the Josephson junction, $E_J$ is the
junction coupling energy, and $E_C$ is the electrostatic Coulomb energy for a single Cooper pair. The parameter of
nonlinearity and driving amplitude for this case reads as
$\chi=-\frac{E_C}{48}$,
$\Omega=-\frac{I\Phi_0}{8\sqrt{2}\pi\eta^{3/2}}$, where $I$ is
the biasing current and $\Phi_0=h/2e$ is the superconducting flux
quantum $\eta=\sqrt{E_J/E_C}$. The typical values for the 
$|0\rangle\rightarrow|1\rangle$ transition frequency in the Josephson junction
used in the experiments cooled to $20 mK$ are $\sim 10 GHz$ and $\sim 10 \mu A$
for biasing currents.
The possible decay path $\gamma/2\pi$ is on the order of
2.6 MHz, and hence, $\chi/\gamma$ is 6.9 \cite{experiment1}, \cite{experiment2}.
Thus, the current-biased JJ combines negligible dissipation with extremely
large nonlinearity.
However, a large nonlinearity in the Josephson inductance is obtained by
biasing the junction at a current $I$ that is very close to the critical
current. As nonlinearity of the Josephson inductance breaks the degeneracy of
the oscillatory energy level spacings, the dynamics of states in near to
resonant excitation is close to the two-level one.

We have included dissipation and decoherence in the NDO on the basis of the
master equation,
\begin{equation}
\frac{d\rho}{dt} =-\frac{i}{\hbar}[H, \rho] +
\sum_{i=1,2}\left( L_{i}\rho
L_{i}^{+}-\frac{1}{2}L_{i}^{+}L_{i}\rho-\frac{1}{2}\rho L_{i}^{+}
L_{i}\right)\label{master},
\end{equation}
where $L_{1}=\sqrt{(N+1)\gamma}a$ and $L_{2}=\sqrt{N\gamma}a^+$ are the
Lindblad operators, $\gamma$ is a dissipation rate, and N denotes the mean
number of quanta of a heath bath.
To study the pure quantum effects we focus on the cases of very low reservoir temperatures which, however, ought to be still
larger than the characteristic temperature $T \gg T_{cr}=\hbar\gamma/k_B$. This
restriction implies that dissipative effects can be described self-consistently
in the frame of the Lindblad Eq. (\ref{master}). For clarity, in our numerical
calculation we choose the mean number of reservoir photons $N=0$. Note, that for
$N\ll1$ the above mentioned restriction is valid for the majority of problems
of quantum optics and, particularly, for the schemes involving the nano-mechanical
oscillator and Josephson junction.
In experiments, the nonlinear oscillator based on the current-biased JJ is cooled down to $T=20mK$ which corresponds to $N=0.0013$,
whereas $T_{cr}=10^{-5}$ K, for $\gamma=1 MHz$.

The time evolution of a NDO driven by a coherent force cannot be solved
analytically for arbitrary evolution times and suitable numerical methods have
to be used. Nevertheless, a dissipative driven nonlinear oscillator has been solved exactly (that means consideration of all order of dissipation) in the steady-state regime and in terms of the
Fokker-Planck equation in complex P representation \cite{i34}, \cite{Drummond}.
An analogous solution has
been obtained for a driven parametric oscillator combined with Kerr
nonlinearity \cite{a33}. The Wigner functions for both these models have been obtained using these solutions \cite{a33}, \cite{i34}. The investigation of quantum dynamics of a driven
dissipative nonlinear oscillator for nonstationary cases, particularly, for
various pulsed regimes, is much more complicated and only a few papers have
been written in this field up to now.

Quantum effects in a NDO with a time-modulated driving force have been studied
in a series of papers \cite{qsch}, \cite{AMK}-\cite{mpop} in the context of a
quantum stochastic resonance \cite{AMK}, quantum dissipative chaos \cite{K2},
\cite{mpop}, and quantum interference assisted by a bistability \cite{qsch}.
The dynamics of the periodically driven nonlinear oscillator also has been
studied in the Refs. \cite{AID}-\cite{RD}. The application of quantum scissors
\cite{LK} to a periodically driven cavity with a Kerr medium has been
considered in Ref. \cite{LM} for the truncation of a coherent state to a
superposition of a vacuum and single-photon Fock state.

\begin{figure}
\includegraphics[width=8.6cm] {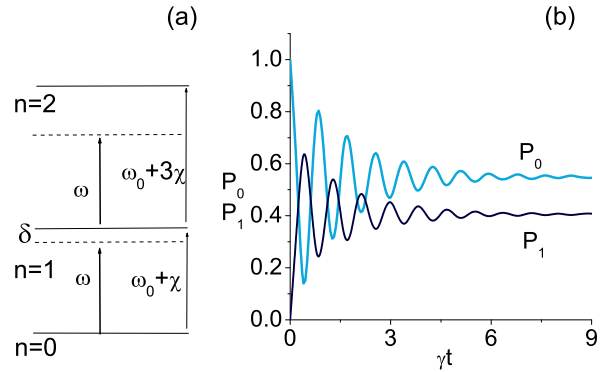}
\caption{(Color online) (a) Energy levels of anharmonic oscillator and (b) Rabi
oscillations of the state populations with decoherence which suppresses beating. The parameters are as follows: $\Delta/\gamma = -11$, $\chi/\gamma=15$, and $\Omega/\gamma = 7$} \label{stat}
\end{figure}

\begin{figure}
\includegraphics[width=8.6cm]{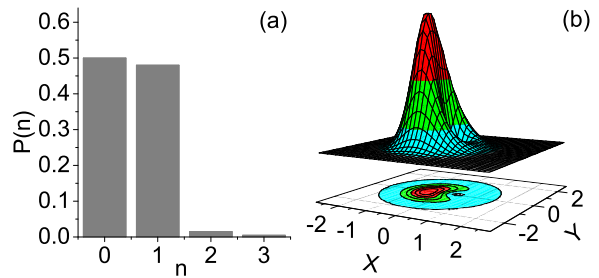}
\caption{(Color online) (a) The excitation numbers distribution and (b) the Wigner function for the NDO in steady-state. The parameters are as in Fig. \ref{stat}.} \label{statW}
\end{figure}

\section{Monochromatically driven NDO: Rabi oscillations and Wigner functions}

In the absence of any
driving, the quantized vibration states of the nonlinear oscillator
are the Fock states $|n\rangle$, which are spaced in energy $E_{n} = E_{0} +
\hbar\omega_{0} n + \hbar\chi n^2$ with $n = 0, 1, ...$. The
levels form an anharmonic ladder [see Fig. \ref{stat}(a)] with
anharmonicity that is given by $E_{21}-E_{10}=2\hbar\chi$. In the case of monochromatic excitation the energetic spectrum is also shifted and in the second-order of
perturbation, the theory with the interaction part $\hbar(\Omega a^{+}
+ \Omega^{*}a)$ of the Hamiltonian described by Eq. (\ref{hamiltonian}) reads as
$\Delta E_{n} = \hbar\Omega^{2} (\frac{n}{\omega + \chi(2n-1)} - \frac{(n
+ 1)}{\omega + \chi(2n+1)})$.

Below we concentrate on quantum regimes
for the parameters when oscillatory energy levels are well
resolved considering near to resonant transitions between lower number
states $|0\rangle\rightarrow|n\rangle$. We solve the master equation Eq.
(\ref{master}) numerically based on the quantum state diffusion method \cite{qsd}.
The applications of this method for studies of the NDO can be found in Refs. \cite{AMK}-\cite{mpop}.
In the calculations, a finite basis of number states $|n\rangle$ is kept large
enough (where $n_{max}$ is typically 50) so that the highest energy states are
never populated appreciably. In the following the distribution of oscillatory
excitation states
$P(n)=\langle n|\rho|n\rangle$ as well as the Wigner functions,
\begin{equation}
W(r, \theta)=\sum_{n,m}\rho_{nm}(t)W_{mn}(r,\theta)
\label{expr:wigner}
\end{equation}
in terms of the
matrix elements $\rho_{nm}=\langle n|\rho|m\rangle$ of the density
operator in the Fock state representation
will be calculated.
Here $(r,\theta)$ are the polar coordinates in the complex
phase-space plane, $x=r\cos\theta$, $y=r\sin\theta$, whereas, the
coefficients $W_{mn}(r,\theta)$ are the Fourier transform of
matrix elements of the Wigner characteristic function.

We start with a discussion of the main idea by considering a first stage NDO
under monochromatic excitation, [$f(t)=1$ in the
Hamiltonian described by Eq. (\ref{hamiltonian})]. It is well known that this system exhibits
regions of regular and bistable motions with the control parameters $\chi,
\Delta, and \Omega$. Below, we consider the monostable operational regime of a pulsed NDO
that is realized for negative values of the detuning satisfying
$\chi(\Delta+\chi)\geq 0$ \cite{Drummond}.

The results for the lowest
excitation $|0\rangle\rightarrow|1\rangle$ are depicted in Fig. \ref{stat} for
the parameters $\Omega/\gamma$ and the detuning
$\delta=\frac{1}{\hbar}(E_1-E_0)-\omega=\Delta + \chi$ meeting the near to resonant
condition, $\delta=4\gamma$. In this case, only two levels effectively are involved in the
Rabi-like oscillations of the populations $P_0$ and $P_1$ of the
vacuum and first excitation number state $|1\rangle$ that is demonstrated in Fig. \ref{stat} (b).
These oscillations take place on the Rabi frequency $R=\sqrt{\Omega^2
+\delta^2}$ that characterizes a two-level quantum system in a monochromatic field, whereas decay of the amplitude takes place due to dissipation. Here, we have 
assumed that the initial oscillatory state is a vacuum state.

Analyzing a monochromatically driven NDO in an over transient regime for time
intervals, $t\gg\gamma^{-1}$, we use both a numerical method and the analytical
results obtained in terms of the exact solution of the Fokker-Planck equation
\cite{a33}, \cite{i34}.
The exact analytical results for a dissipative-driven nonlinear oscillator have
been obtained in terms of the Fokker-Planck equation in the complex
P-representation. In this way, the solution for the Wigner function of the 
oscillatory mode involving quantum noise in all order of perturbation theory
and in the steady-state regime have been derived
\begin{equation}
W(\alpha)=N\exp^{-2|\alpha|^{2}}\left|\frac{J_{\lambda-1}(\sqrt{-8\alpha\varepsilon})}{(\alpha^{*})^{(\lambda-1)/2}}\right|^{2}.
\end{equation}
Here, $J_{\lambda}$ is the Bessel function,
$\lambda=(\gamma+i\Delta)/\chi$, $\varepsilon=\Omega/\chi$, and
amplitude $\alpha=x+iy$ is the complex C-number variable corresponding to
the operator $a$. It is evident from this result that, under monochromatic force, the
NDO does not display any quantum interference pattern and cannot generate superposition states as the corresponding
Wigner function is positive in all phase space. It should be noted that the
steady-state solution of the Fokker-Planck equation has been found using the
approximation method of potential equations \cite{Drummond}, \cite{Haken}. The validity of
this solution has not been checked up to now in the strong quantum regime that
requires a high nonlinearity with respect to dissipation. For this reason, we also calculate the Wigner function numerically on the basis of the numerical simulation of the 
master equation by using the quantum state diffusion method \cite{qsd}. According
to these calculations, the system reaches the equilibrium of oscillatory
states in the steady-state regime and the Wigner function of the oscillatory mode is
positive in all phase-space describing statistical mixtures of states
$|0\rangle$ and $|1\rangle$. The result is displayed in Fig. \ref{statW} for
the parameters that have been used in the Fig. \ref{stat}.

\section{Production of Fock states in the pulsed regime}

Now, we are able to present the main results of this paper
concerning production of Fock states and qubits for the NDO due to
pulsed excitation. In this sense we note the main
peculiarity of our paper. We investigate the production of
nonclassical states for interaction time intervals
exceeding the characteristic time of the dissipative processes, $t\gg\gamma^{-1}$, however, in the nonstationary regime that is
conditioned by the specific form of excitation.

\begin{figure}
\includegraphics[width=8.6cm]{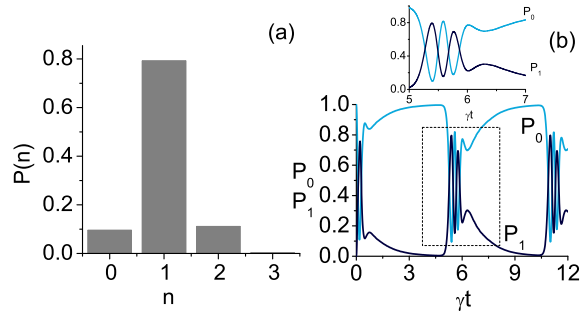}
\caption{(Color online) The excitation number distribution for time intervals corresponding to
the maximal population (a) of the $|1 \rangle$ state and (b) time-dependent populations. The inset shows the enlarged graphic to visualize Rabi-type oscillations in detail. The parameters are as follows:
$\Delta/\gamma= -15$, $\chi /\gamma = 15$, $\Omega/\gamma = 6$,
$\tau = 5.5{\gamma}^{-1}$, and $T=0.4{\gamma}^{-1}$.}
\label{fed}
\end{figure}

In this way, we consider two important regimes leading to the production of both Fock states and superposition of Fock states for the low-power
resonance transition $|0\rangle\rightarrow|1\rangle$ between vibrational
states. Each of these regimes is realized for the appropriate choosing
parameters, the detuning $\Delta$, the Rabi amplitude $R$, and the
parameters of pulses. For the pulsed NDO, the ensemble-averaged mean
oscillatory excitation numbers, the populations of oscillatory
states and the Wigner functions are nonstationary and exhibit a
periodic time dependent behavior, i.e., repeat the
periodicity of the pump laser in an over transient regime.
\begin{figure}
\includegraphics[width=8.6cm]{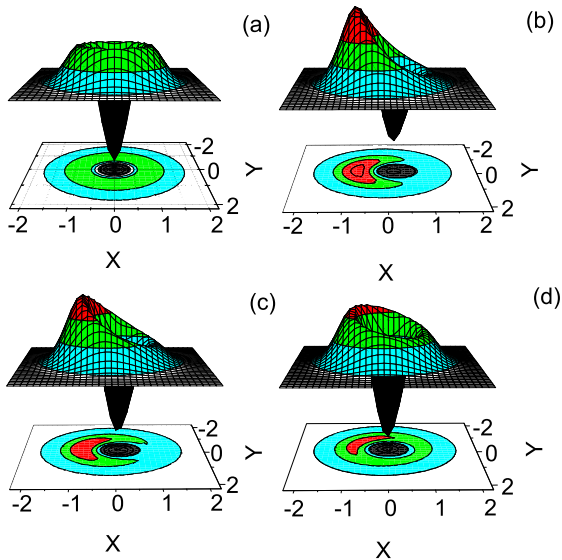}
\caption{(Color online) (a) The Wigner function for pure $|1\rangle$ state. 
Evolution of the Wigner function for (b) $t=3\tau - 0.5T$, (c)
$t=3\tau - 0.4T$, and (d) $t=3\tau-0.25T$ for the parameters:
$\Delta/\gamma=-15$, $\chi /\gamma = 15$, $\Omega/\gamma = 6$, $T=
0.4{\gamma}^{-1}$, and $\tau =5.5{\gamma}^{-1}$. The ranges of negativity are indicated in black.}
\label{fwig}
\end{figure}

We demonstrate below that Fock state $|1\rangle$ can be effectively
produced if the pulsed excitation is tuned to the exact resonance $\delta=\Delta +
\chi=0$. The other condition concerns the controlling parameters of the
pulses. We assume the regime involving short pulses separated by long time
intervals that allows increasing the weight of a single Fock state. In these
cases, the probability distribution of excitation number $P(n)$ displays a maximum approximately equal to unity for the one-photon state $n = 1$. The condition
of low excitation reads as $|\Delta/\Omega| > 1$. It allows avoiding mixed
states, as the external field populates both $|n+1\rangle$ and $|n-1\rangle$
states. The typical results for pulse duration $T=0.4\gamma^{-1}$ and the time
interval between pulses $\tau = 5.5\gamma^{-1}$ are shown in Fig. \ref{fed}.
The time evolution of the
probabilities $P_0$ and $P_1$ of the $|0\rangle$ and $|1\rangle$ states starting
from the vacuum state are depicted in Fig. \ref{fed}(b). As we see, the maximal weight
0.8 of $P_1$ is realized for the definite time-intervals of measurement $t =
k\tau - 0.25T$, $k = 1, 2, 3,...$, [see Fig. \ref{fed}(a)].

Note that, due to the anharmonicity, this system is close to the two-level
limit for small amplitudes, and the resonant excitation and the time evolution
are essentially quantum. As is well known, in the case of a resonant
excitation of a two-level atomic system by a laser pulse with the electric field
envelope $E(t)$, the population of the excited state oscillates as
$sin^2(\frac{1}{2}R(t))$, where $R(t)=\frac{1}{\hbar}\int_{-\infty}^{t}{dt'(E(t')d)}$ is the time dependent Rabi frequency,
and $d$ is the dipole moment of atomic transition.
For the resonant excitation of the nonlinear oscillator under a single pulse, the Rabi frequency is determined by the time-dependent coupling constant $\Omega(t)$ (see also Ref. \cite{claud}).
In this case, the Rabi-like oscillations have been observed for the Josephson
device driven by resonant microwave flux pulses that behaves as an anharmonic
oscillator \cite{claud}. The corresponding Wigner functions that show negative
values have been also calculated in decoherence free approximation for time
intervals during the Rabi period.
If the NDO is driven by a sequence of pulses, this behavior
is modified and essentially depends on the duration of the pulses as well as
from time intervals between them.
As our calculations show, in this case, Rabi-like oscillations take place for
short time intervals between two adjacent pulses. The detailed description of
such oscillations is depicted in Fig. \ref{fed}(b), the inset graphic.
Analyzing time evaluation of the Wigner functions during pulses we
conclude the production of Fock state $|1\rangle$ in the pulsed
NDO. The results of the calculations of the Wigner function are presented
in Fig. \ref{fwig} for various time intervals within the pulse duration and are
compared with the Wigner function of the pure state $|1\rangle$ [see Fig. \ref{fwig}(a)]. It is easy to realize that the Wigner function depicted in Fig. \ref{fwig}(d) for time interval $t=3\tau - 0.5T$
corresponding to the maximal population $P_1$ displays a ring signature with the
center at $x = y = 0$ in the phase-space that is a characterized pure Fock state
and, hence, is closest to the pure state $|1\rangle$ Wigner function [see Fig.
\ref{fwig}(a)]. An intuitive explanation of this point has been performed above on the basis of the results depicted in Fig. \ref{fed}.

\section{Quantum superposition of Fock states}

\begin{figure}
\includegraphics[width=8.6cm]{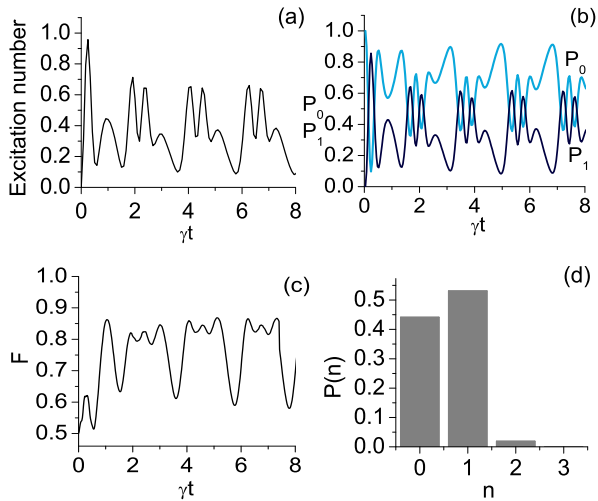}
\caption{(Color online) (a) The mean excitation number, (b) populations of
oscillatory states, (c) time evaluation of the fidelity, and (d) the
distribution of excitation number for time interval t = $5\tau -
0.4T$. The parameters are as follows: $\Delta/\gamma = -11$, $\chi
/\gamma = 15$, $\Omega/\gamma = 7$, $T = 0.7{\gamma}^{-1}$, and $\tau
= 2.2{\gamma}^{-1}$.} \label{sef}
\end{figure}

\begin{figure}
\includegraphics[width=8.6cm]{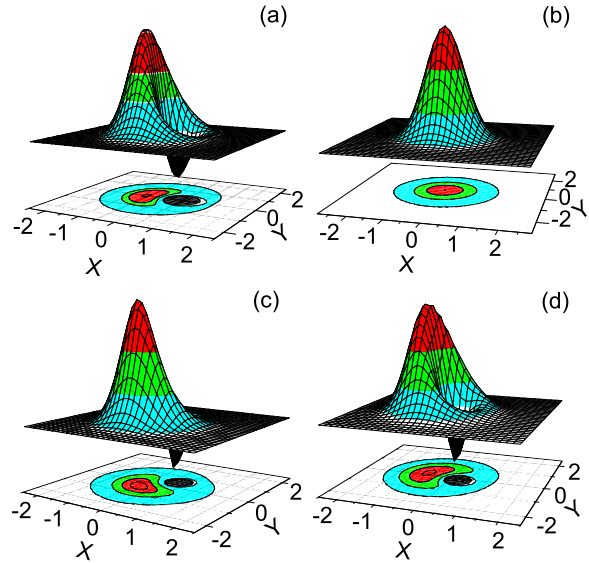}
\caption{(Color online) (a) The Wigner function for the pure $\frac{1}{\sqrt{2}}(|0\rangle - |1\rangle)$ state. The Wigner function evaluation for the parameters used in Fig. \ref{sef}: (b) $t=5\tau -1.4T$, (c) $t=5\tau - 0.9T$, and (d) $t=5\tau - 0.4T$. The ranges of negativity are indicated in black.}
\label{fig3}
\end{figure}

The most striking signature of the periodically pulsed NDO is the appearance of quantum interference between Fock
states in the over transient regime. We demonstrate it for near-to-resonant transition $|0\rangle\rightarrow|1\rangle$ with the oscillatory parameters used above in Figs. \ref{stat} and \ref{statW}.
The typical results for the dynamics of the excitation numbers and the Wigner
function for the pulse duration $T = 0.7\gamma^{-1}$, and the time interval
between pulses $\tau = 2.2\gamma^{-1}$ are demonstrated in Figs. \ref{sef} and
\ref{fig3}.

A qualitative explanation for the production of Fock superposition states can be
obtained by considering the time evolution of the anharmonic oscillator in the overtransient regime, where dissipation and decoherence effects are essential. This
analysis allows us to choose the time intervals for the effective joint excitation
of states $|0\rangle$ and $|1\rangle$. In this way, time evolution of the mean
excitation number and the populations of vacuum and excited states are
calculated and are depicted in Figs. \ref{sef}(a) and \ref{sef}(b).
As we see, for the overtransient regime, the time modulation of the averaged excitation number and the populations of two oscillatory states repeat the periodicity of the pump laser. The populations display time dependent Rabi-like oscillations in the ranges of time intervals between two adjacent pulses that offer a clear signature of quantum behavior and the transitions between the lowest eigenstates.
The existence of a Rabi-like oscillation under the influence of the driving, which is resonant with transitions between neighboring levels, is closely linked to the
anharmonicity of the oscillator so that oscillations do not show up in a
harmonic oscillator driven by resonant monochromatic excitation in the overtransient regime.
Moreover, it is intuitively clear that quantum interferences and the forming of
Fock superposition states are possible in the course of these Rabi-like
oscillations that involve the crossing of two populations.

To demonstrate this statement, we investigate the Wigner function for the
parameters using in Fig. \ref{sef}. Using these data, in Fig. \ref{fig3}, we plot the evolution of the Wigner function that demonstrates interference
fringes on the phase space between $|0\rangle$ and $|1\rangle$ states. Analyzing
this evolution, we conclude that the closest to the pure superposition state
$|\Psi\rangle=\frac{1}{\sqrt{2}}(|0\rangle-|1\rangle)$, the Wigner function [see,
Fig. \ref{fig3}(a)] is realized for the measurement time intervals $t=k\tau-0.5T$,
k = 1, 2, 3,... within the regime of Rabi-like oscillations for which the
populations of the two lower states are approximately equal [see Figs. \ref{sef}(d)]. This conclusion is also supported by the calculation of the time-dependent fidelity
$F=|\langle\Psi|\rho|\Psi\rangle|$ depicted on Fig. \ref{sef}(c). As we see,
the time-dependence of the fidelity repeats the evolution of the pulses and
reaches the maximal values 0.88 for time intervals $t=k\tau-0.5T$, k = 1, 2,
3,..., corresponding to the Wigner function in Fig. \ref{fig3}(d). It is important that the maximal fidelity is realized when the oscillatory mean excitation
number reaches its maximal value, thus quantum interference effects between
$|0\rangle$ and $|1\rangle$ states are strong in the case of a high power
excitation.
We realize that production of quantum interference in the overtransient regime is
due to the control of the decoherence through the application of suitable
tailored, synchronized pulses [see, for example, Refs. \cite{kicked}, \cite{ent}]. Indeed, quantum interference is realized here if a mutual influence of pulses
is essential (for $\tau/T = 3.14$ in Fig. \ref{fig3}). In this case the decay
of Rabi-like oscillations is negligible for time intervals between the pulses
that is reflected in the production of quantum interference for the dissipative
regime.

\section{Conclusion}

We have shown that the production of the Fock states as well as the
superposition of the Fock states are possible in the single-mode dissipative
oscillator due to strong Kerr nonlinearities and the periodically pulsed
excitation. The nonlinearities should be strong enough to avoid the degeneracy
of the oscillatory energy-level spacings as well as to reach quantum
operational regimes, that are realized for $\chi/\gamma>1$. Quantum
interference between Fock states in the overtransient regime is realized due to
driving the oscillator by a series of short pulses with proper parameters for
effective reducing of the dissipative and decoherence effects. These results
have been demonstrated by numerical calculations of the dynamics of oscillatory
exsitation numbers and the Wigner functions of oscillatory mode for the various
over transient regimes of NDO. We hope that this approach can be expanded to
include the production of the other lower Fock states $|2\rangle$, $|3\rangle$,
etc.

This problem of engineering various quantum states has attracted increasing
interest in recent years not only for quantum information processing, but also
for various other quantum technologies based on nano-mechanical oscillators,
particularly, in optomechanics, for the manipulation of mechanical degrees of
freedom by light scattering. We believe that the scheme under consideration,
operated with a train of realistic Gaussian pulses, is available in these
areas for experiments with laser or microwave pulses [see, for example
Refs. \cite{claud} and \cite{Tian}]. Together with the recent advancements in
the engineering of various cooled oscillatory systems, these results seem to be
important for the further studies of quantum phenomena in light of new
technologies.

\end{document}